# Terahertz composite plasmonic slabs based on double-layer metallic gratings

**DEJUN LIU,**[1,*] **LIN CHEN,**[2] **AND FENG LIU,**[1]

[1]*Depatment of Physics, Shanghai Normal University, Shanghai 200234, China*
[2]*Shanghai Key Lab of Modern Optical System, University of Shanghai for Science and Technology, Shanghai 200093, China*
*\*dejunliu1990@gmail.com*

**Abstract:** A composite plasmonic slab based on double-layer metallic gratings and a dielectric film is experimentally and numerically demonstrated in terahertz (THz) regions, which can support two resonance modes and then form a broad bandgap (40%). As compared to the double-layer metal grating, the dielectric film in composite THz slabs significantly enhances the transmission of the transverse magnetic (TM) mode. Electric field vector proved that the low-frequency resonance mode originates from the symmetric plasmonic mode and the high-frequency resonance mode is induced by the hybrid mode of plasmonic and dielectric modes. The inherently near field coupling between metal gratings and dielectric film has been analyzed by changing the structural parameters. We further demonstrate that by tuning the metallic grating width, the plasmonic bandgap can be manipulated. These results suggest that this composite plasmonic slab has great potential for use as a filter, polarizer, and sensor in THz regions.

## 1. Introduction

Terahertz (THz) radiation, defined as the frequency of 0.1-10 THz, arouses increasing interest for important applications in a wide range of fields, such as imaging [1], sensing [2], and spectroscopy [3]. A device that manipulating the THz waves is highly desired in these applications. By structuring metal surface, a THz plasmonic structure supports surface plasmon polaritons (SPPs) that realizes a high confinement mode. Tremendous efforts have been devoted to realizing multi-functional plasmonic devices, including changing the geometrical shapes, the interval between the structure units, and the structure stacks. Typical plasmonic structures such as metallic hole and slit arrays with extraordinary transmission properties have been extensively attempted for this purpose because they can concentrate optical energy in a nanoscale spatial region and allow for tailoring the spectra response [4-11]. For instance, a metal hole array with a thin dielectric layer was attempted for a bandpass property, where the field coupling between the metallic structure and the dielectric film determines the spectral response and resonance frequencies [12]. In order to improve the coupling effect, two-layered metal stacked structures have been proposed [13-16]. Double-layer metal hole arrays with unexpected transmission characteristics have been found by varying the layer spacing and the lateral displacement, where the near-field coupling of surface wave plays a key role [13]. Unfortunately, two separated metal layers generally suffer from a giant loss and need precise control. Therefore, a sandwich structure consisting of double-layer metal slit arrays with a silicon film spacer has been reported, where two layers of the metal slit are close proximity [16]. The later shifts of two metal layers have giant effect on the peak transmission. The smallest peak occurs at lateral shifts of a quarter of the grating period. A dielectric film with a plasmonic structure forms novel structures call hybrid plasmonic structures. Hybrid plasmonic structures (HPS) are becoming attractive since their remarkable abilities were found to improve the trade-off between the confinement and attenuation of the conventional plasmonic devices [17]. An HPS consists of a dielectric and metal surfaces, which enable adjustment of the coupling efficiency between dielectric mode and SPP modes [18-23]. However, the dielectric film contributes to the hybrid mode has not been well analyzed in these multi-layer structures [16-32]. In addition, little analysis of the coupling of the fields in multi-layer hybrid plasmonic slabs has been done on the THz frequency range.

Here we proposed a composite plasmonic slab based on double-layer metallic gratings and a dielectric film at THz frequencies. This slab exhibits two resonant peaks, corresponding to the symmetry plasmonic and hybrid modes. The experimental results agree well with simulation, and the calculated field profiles confirm the physical origins of the two transmission peaks. The hybrid mode originates from the mixing of plasmonic and dielectric modes, which has been analyzed by the electric field vector and power distribution. We also investigate the effect of the structural parameters such as the thickness and refractive index of the dielectric film between the double-layer metal gratings on the two resonant peaks. Furthermore, the influence of the grating width on the plasmonic bandgap has been well discussed. Our works analyzed the role of the dielectric film and near-field coupling in the hybrid slab, which is attractive for the design of the optoelectronic device in the THz gap.

## 2. The composite plasmonic slabs

The proposed composite plasmonic slab (CPS) consists of double-layered metal gratings and a dielectric film. The schematic diagram of this structure with all geometric parameters is illustrated in Fig. 1. The structural parameters are as follows: $\Lambda$=1.5 mm, w=0.5 mm, s=1.0 mm, t=0.2 mm. The upper and lower metal gratings were isolated by a dielectric spacer of 0.2 mm-thick polyethylene terephthalate (PET) with refractive index of 1.6 and loss tangent ($\delta$) of 0.0442 in THz regions [33]. The metallic layers on opposing sides of the substrate film are copper with a thickness of 0.1 mm. The total thickness of the structure is 0.4 mm, which is smaller than the wavelength (sub-wavelength). The image of the experimental sample of a single-layer metal grating is given in Fig.1. The composite plasmonic slab in the measurement is a rough sample, which is assembled by hand. Consequently, the alignment between the top and bottom metal gratings is not good. The transverse magnetic (TM) polarization is defined as a polarization state, where the electric field orientation is perpendicular to the metal grating. In the simulation, we apply periodic boundary conditions in the X- and Y- directions while open condition along the Z-direction. Simulations of the structures are mainly achieved by the finite element method (FEM) [34].

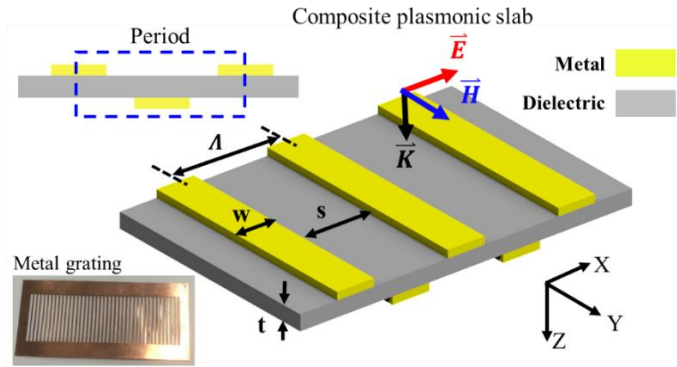

Fig. 1. Sketch of the composite plasmonic slab. The slab consists of double-layer metal gratings and a dielectric film.

## 3. The results and discussion

### 3.1 Experimental and simulated results

The experimental results are measured using the THz-TDS system (Advantest 7400). A signal of air speace was used as a reference. Fig. 2 (a) shows the THz waveform of reference, single-layer metal grating, and composite plasmonic slab. The time-domain signal is attenuated by the metal grating due to the rough metal edges. The THz signal shows clear attenuating oscillation after passing through the double-layer metal gratings combining with a PET film, indicating a strong interaction between the THz waves and the formed composite slab. Their frequency-domain spectra obtained by Fourier transformation are shown in Fig. 2 (b). A resonant peak is occurred at 0.184 THz, resulting from the fundamental mode of metal gratings. With the increase of frequency, the transmission power is obvious decay. The blue dot line is the experimental spectrum of the composite plasmonic slab (CPS). The resonant peak is 0.166 THz, which shows a redshift comparing with single-layer metal grating. Interestingly, a resonant peak with high transmission power occurs at 0.312 THz. The simulation results (grating line) agree well with those experiment results. For example, the second resonance peak is 0.332 THz in the simulation transmission spectrum. Note that the peak frequency and amplitude discrepancy between the experiment and the simulation spectra for the CPS are possibly due to imperfections in the sample's combination process. Two resonant peaks are separated by a bandgap with a bandwidth of 0.102 THz. These results indicate that this CPS can be used as THz filters.

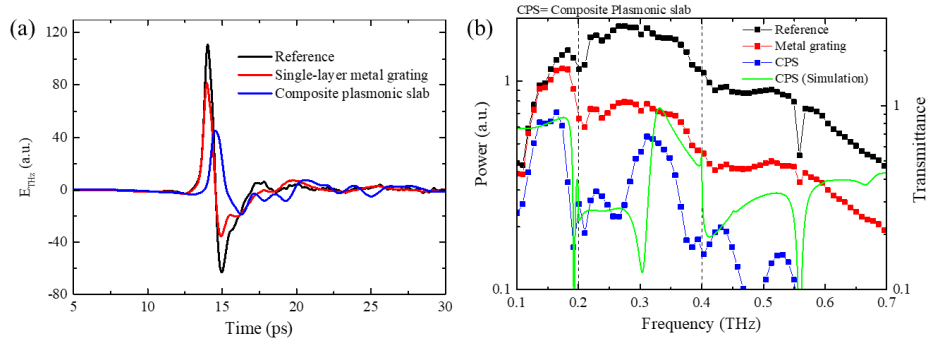

Fig. 2. (a) THz waveform and (b) transmission spectra of reference, single-layer metal grating, and composite plasmonic slab (CPS). The green line in Fig. 2 (b) is the simulation result.

*3.2 Propagation modes in composite plasmonic slabs*

To better understand the composite slab, we will analyze the propagation mode of TM waves in metal gratings. We suppose that TM waves propagate in the +z-direction. The cut-off frequency of propagation modes can be expressed as follows [35]:

$$f_c = \frac{cm}{2sn_{eff}}$$

where $c$, $m$, and $n_{eff}$ are light speed in vacuum, the number of waveguide modes, and effective waveguide refractive index, respectively. For the case of s=1.0 mm, the fundamental TM mode ($m=0$) has no cut-off frequency. In contrast, the $TM_1$ mode ($m=1$) has a cut-off frequency of 0.2 THz (the effective refractive index $n_{eff}=1$) and the $TM_2$ mode has a cut-off frequency of 0.4 THz. The simulated spectra of metal gratings for TM waves are depicted in Fig. 3 (a) (black line). For the metal gratings, the obvious decreases at 0.2 and 0.4 THz can be observed (the blue dot line). Based on the theory of waveguides, 0.2 and 0.4 THz respectively corresponds to the cut-off frequency of $TM_1$ and $TM_2$ modes. Therefore, the frequency range from 0.2 to 0.4 THz is the mixing modes of $TM_0$, $TM_1$, and $TM_2$. When the metal grating integrated with a dielectric film, one resonance peak at 0.345 THz is occurs. It means that the resonance mode is enhanced by the dielectric film. Figures 3 (c) and (d) show the electric field distribution for single-layer metal grating with a dielectric film at 0.195 THz and 0.345 THz, respectively. Apparently, the strong electric field is confined in the surface regions of the dielectric film and metal gratings. It is noted that the dielectric mode combining with local fields improves the field confinement and thus causes the high transmittance at 0.345 THz.

Figure 3 (b) shows the simulated transmittance spectra of double-layer gratings (black line) and composite slabs consisting of double-layer gratings with a dielectric film (red line). The double-layer metal grating structure behaves as a Fabry-Perot (FP) cavity where multiple reflections occur at two interfaces. Furthermore, the SPP's are excited on double-layer metal structures couple with each other and consequently resulted in a change in the transmission spectrum. The resonance mode at 0.195 THz is enhanced by the double-layer metal grating structure. A similar phenomenon can be observed in the reference [13]. In the region of 0.2–0.7 THz, the double-layer metal gratings exhibit a lower transmission over the frequency region of interest compared to the single-layer metal grating. The changes in the transmission spectrum come from the near-field coupling between the two surface waves excited on the double-layer metal grating. Resonance modes can be enhanced by using the dielectric film. A dielectric film is embedded inside the double-layer metal gratings, which form a composite plasmonic slab. As shown in Fig. 3 (b) (red line), the peak transmittance at 0.332 THz increases from 0.23 to 0.98, which is significantly improved by the dielectric film. We note that there is a slight difference in the frequency of the lower peak transmission for double-layer metal gratings and CPS, which is probably attributed to the coupling between SPP, FP resonance, and dielectric modes.

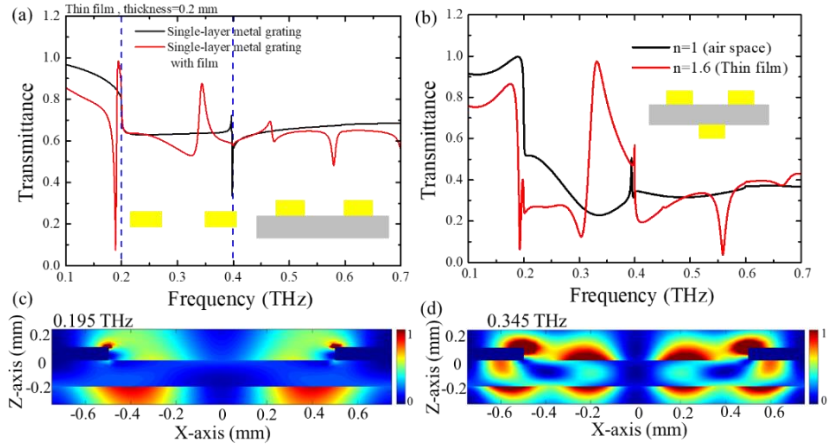

Fig.3. (a) The transmission spectra of single-layer metal grating and single-layer metal grating with a dielectric film. (b) The spectra of double-layer metal gratings with air space and a dielectric film. (c-d) The electric field distribution of single-layer metal grating with a dielectric film at 0.176 THz and 0.345 THz.

Figure 4 (a) shows the transmission spectra of the composite slabs. Two noticeable transmission peaks with high transmittance (>80%) at 0.176 and 0.332 THz can be observed, respectively. Two peaks in the spectrum are separated by a broad bandgap from 0.198 THz to 0.3 THz, where a 0.102 THz (40%) bandgap centered in the vicinity of 0.25 THz. This bandgap is broader than the previous studies of photonic crystals [36-37]. By changing the structure period and metal width, the bandgap frequency and spectral contrast can be manipulated. For the case of 1.0 mm-$\Lambda$ and 0.4 mm-w, a bandgap with a bandwidth of 0.15 THz is achieves. Electric field distributions of 0.176 and 0.332 THz for the 1.5 mm-$\Lambda$ structure are demonstrated in Figs. 4 (b) and (c). The localized electric field of 0.176 THz is mainly confined at the edge of metal gratings, indicating that this resonance mode is induced by the SPPs and thus can be termed as a plasmonic mode. However, the field pattern at the 0.332 THz behaves as a hybrid mode of the mixing of plasmonic and dielectric modes, where the strong field locates at the edge of metal gratings and dielectric surfaces.

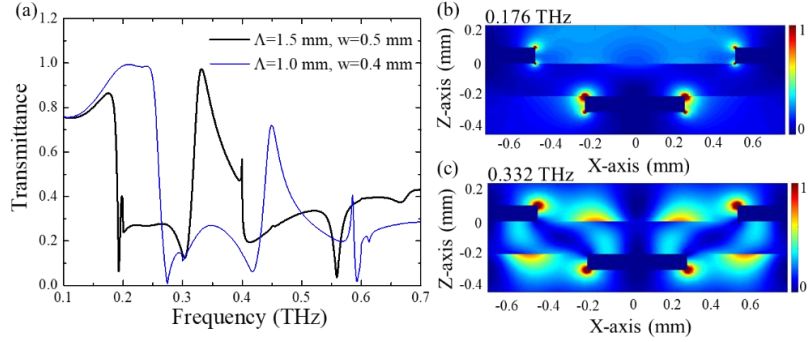

Fig.4. (a) The transmission spectrum of hybrid plasmonic waveguides with a different metal period and width. (b-c) The electric field distribution of 0.176 THz and 0.332 THz in the transmission spectral peaks.

We also simulated the electric field vector and power distribution of 0.176 and 0.332 THz for the composite plasmonic slab. As presented in Fig. 5, the plasmonic mode at 0.176 THz behaves as a symmetric mode with two symmetric parallel dipoles, which coincides with the symmetric mode or anti-bonding mode that has been observed in previous studies [14-15]. On the other side, the mode at 0.332 THz shows asymmetric parallel dipoles, which is different from that of 0.176 THz. At this frequency, the electric displacement at the top and bottom metal grating are opposite to each other and forms a loop. We consider the mode at 0.332 THz is a hybrid mode, where the dielectric mode is dominant. Power distributions are illustrated in Figs. 5 (c) and (d). The intense power of 0.176 THz is located at the center of the gap between metal gratings. Inversely, the intense power of 0.332 THz is confined at the gap regions between metal gratings

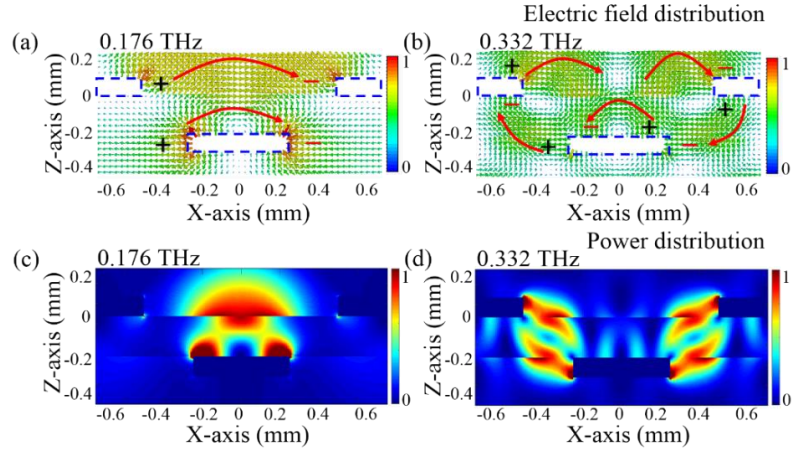

Fig.5. The electric field vector distribution at 0.176 THz (a) and 0.332 THz (b). The power distribution of 0.176 THz (c) and 0.332 THz (d) at transmission spectral peaks.

### 3.3 Resonance modes dependent on the structural parameters

### 3.3.1 The film thickness dependent on the resonance modes

The near-field coupling between metal gratings and dielectric film can be controlled by changing the structural parameters. Since the proposed slab consisting of double-layer gratings and the thickness of the dielectric film is far less than the operation wavelength, a strong near-field coupling effect exists between the two gratings. It can be divided into vertical and lateral couplings, the first one is mainly determined by the dielectric film thickness, while the second one not only depends on the width of metal gratings, but also indirectly relies on the dielectric film thickness [38]. In this part, we investigate the vertical coupling effect in this composite slab. We simulated the transmission spectra and electric field distribution by varying the dielectric film thickness from 0.10 to 0.30 mm, where the refractive index of the dielectric film is fixed as 1.6. As shown in Fig. 6, the spectral responses such as transmittance and peak frequency exhibit different trends with the film thickness increases. The spectral peaks shift to lower frequencies as the film thickness increases. However, the transmittance of low-frequency band peak increases from 0.74 to 0.99 when the film thickness increases from 0.10 to 0.30 mm. It means that resonance mode at the low transmission band is improved by the thick dielectric film. Contrarily, the transmittance of the high-frequency band peak first increases and then decreases. The hybrid plasmonic mode is first enhanced when the film thickness is increasing from 0.10 to 0.20 mm. When the film thickness larger than 0.2 mm, the coupling between spoof SPPs with dielectric modes becomes weaker and thus results in a lower transmittance. It suggests that if the upper grating is far enough from the lower one, no coupling effect exists. Therefore, the vertical coupling effect in this hybrid slab can be tailored by the thickness of the dielectric film. Electric field distributions of 0.362 and 0.308 THz for 0.10 and 0.30 mm are shown in Fig. 6 (c), respectively, where color bar magnitude is normalized to the same value. Obviously, the thin film shows a strong coupling, in which the strong field is located at the surface of the dielectric film and the edge of the grating. When the film thickness is changed to 0.30 mm, the field becomes weaker compared with that of 0.10 mm. The vertical coupling strength can be optimized by changing the film thickness [28]. The film thickness dependent resonance mode shift provides an additional degree of freedom for tuning the resonance frequency.

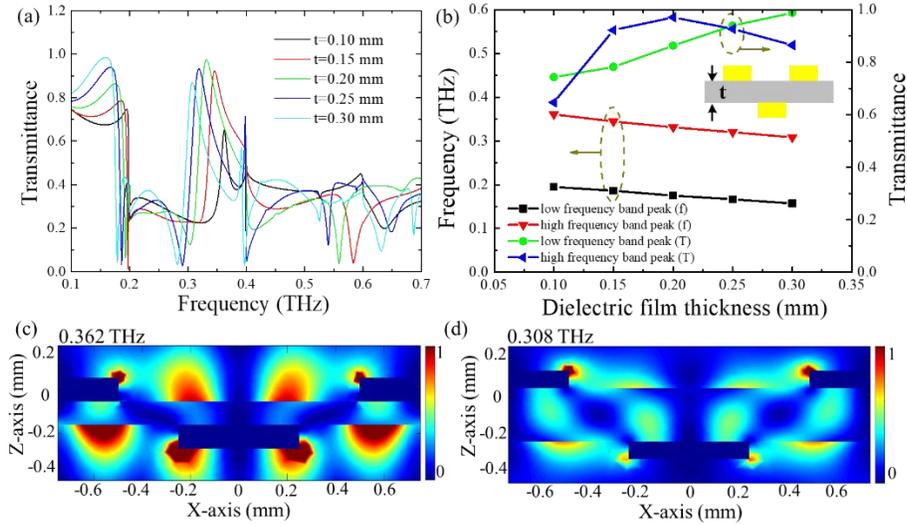

Fig.6. (a) The transmission spectra of composite plasmonic slabs for various dielectric films with different thicknesses. (b) The spectral peak frequency and transmittance for various dielectric films with different thicknesses. (c) The electric field distribution of 0.362 THz and 0.308 THz for 0.10- and 0.30-mm dielectric films, respectively.

*3.3.2 The refractive index of dielectric film dependent on the resonance modes*

Figure 7 shows the spectra and field distribution of composite plasmonic slabs. Dielectric films have different refractive indices, but the thickness is fixed as 0.20 mm. With the film refractive indices increasing, the spectral peaks shift to lower frequencies. The bandwidth of high-frequency mode and bandgap become narrower, which also can be observed from the contour map in Fig. 7 (b). An enhanced electric field at 0.373 THz is achieved by the dielectric film with a low refractive index (n=1.2), which tightly confined by dielectric film and metal gratings. The field becomes weaker when the film's refractive index altered to 2.0. It means that the vertical coupling effect relies on the refractive index of the dielectric film.

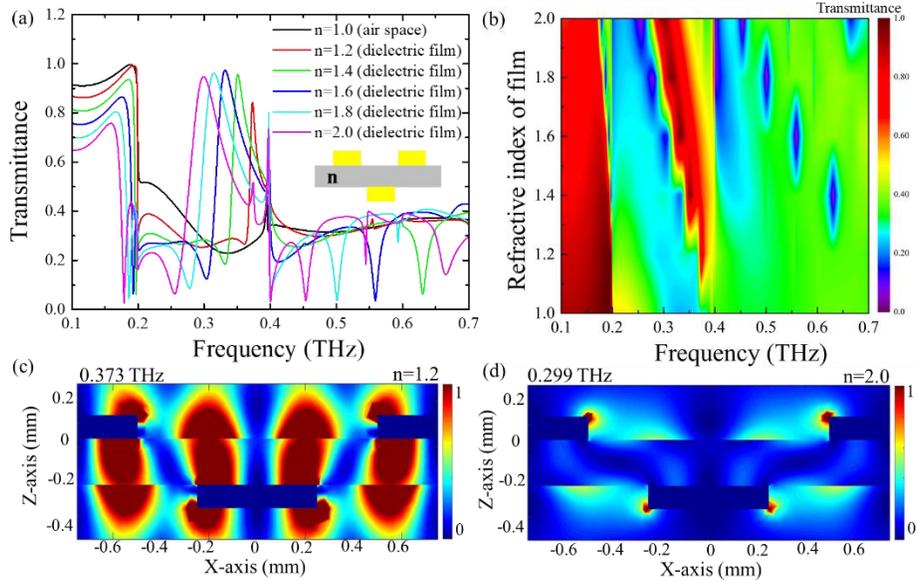

Fig.7. (a) The transmission spectrum of composite plasmonic slabs for various dielectric film refractive indices. (b) The contour map for various dielectric film refractive indices. (c-d) The electric field distribution of 0.373 THz and 0.299 THz for n=1.2 film and n=2.0 film, respectively.

*3.3.3 The width of metal gratings dependent on the resonance modes*

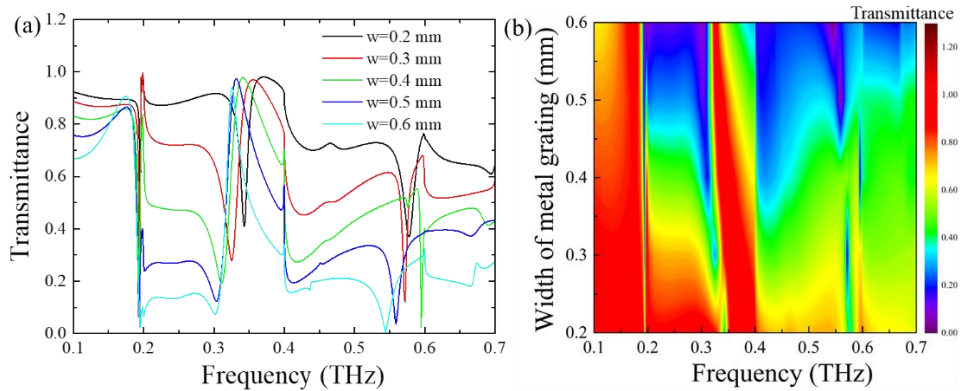

Fig. 8. (a) The transmission spectrum of composite plasmonic slabs for various grating widths. (b) The contour map of composite plasmonic slabs for various grating widths.

We further performed the simulation for the various widths of metal grating when the dielectric film thickness is 0.20 mm and the refractive index is 1.6. The results are illustrated in Fig. 8. The transmission band from 0.197 to 0.34 THz shows a high transmittance when the grating width is 0.20 mm. With the increase of grating width, the obvious bandgap is occurs, where the transmittance at 0.3 THz drops from 0.9 to 0.1. As shown in Fig. 8 (b), results indicate that the smaller grating width achieves higher transmittance due to the weak coupling in the CPS. For the hybrid mode at high frequency band, its bandwidth becomes narrower as the increases of metal width. In other words, the quality (Q) of hybrid modes is improved. It is interesting that only two bands with high transmittance exist when the metal grating width larger than 0.50 mm, suggesting that higher order mode cannot propagate through this composite slab.

## 4. Conclusions

In this work, we propose a composite plasmonic waveguide based on double-layers metallic gratings and a dielectric film, which can support two resonance modes and a broad bandgap (40%). The experimental results agree well with that of simulation. Electric field vector demonstrated that the low-frequency resonance mode at 0.176 THz is a plasmonic mode with symmetric dipole alignment, and the high-frequency resonance mode at 0.332 THz belongs to a hybrid mode that originates from the mixing of plasmonic and dielectric modes. We also investigated the influence of other parameters such as the thickness and refractive index of the dielectric film between the two metal layers on the two resonant peaks. The spectral response of this structure can be efficiently tailored by these structural parameters because of the near-field coupling. The broad bandgap ranges from 0.198 THz to 0.30 THz, which is giant influenced by the metal grating width. We believe that this composite plasmonic slab is attractive for the design of the optoelectronic device in the THz gap and higher frequency regions.

## Funding

This study received funding from the Science and Technology Commission of Shanghai Municipality (No. 19590746000, No. 18590780100 and No. 17142200100), Innovation Program of Shanghai Municipal Education Commission (No. 2019-01-07-00-02-E00032).

## Disclosures

The authors declare no conflicts of interest.